\documentclass[a4paper,fleqn]{cas-dc}
\renewcommand{\printorcid}{}

\usepackage{apacite}
\usepackage[round]{natbib} 

\usepackage{breakcites}

\usepackage{float}
\usepackage[section]{placeins}

\restylefloat{table}
\usepackage{diagbox}
\usepackage{adjustbox}

\usepackage{textgreek}

\usepackage{geometry}

\usepackage[dvipsnames]{xcolor}

\def\tsc#1{\csdef{#1}{\textsc{\lowercase{#1}}\xspace}}
\tsc{WGM}
\tsc{QE}
\tsc{EP}
\tsc{PMS}

\tsc{BEC}
\tsc{DE}

\begin{document}

\let\WriteBookmarks\relax
\def\floatpagepagefraction{1}
\def\textpagefraction{.001}
\shorttitle{Macroeconomic forecasting with statistically validated knowledge graphs}
\shortauthors{Sonja Tilly et~al.}

\pagenumbering{arabic}

\title [mode = title]{Macroeconomic forecasting with statistically validated knowledge graphs}                      

\author[1]{Sonja Tilly}

\cormark[1]

\ead{sonja.tilly.19@ucl.ac.uk}

\cortext[cor1]{Corresponding author}

\address[1]{UCL, Computer Science Dep, 66 - 72 Gower St, Bloomsbury, WC1E 6EA London, UK}

\author[1,2]{Giacomo Livan}

\address[2]{Systemic Risk Centre, London School of Economics and Political Science, London, WC2A 2AE, UK}
\ead{g.livan@ucl.ac.uk}

\nonumnote{Abbreviations. GDELT: Global Database of Events, Language and Tone; GKG: Global Knowledge Graph; CNT: Conviction Narrative Theory; Bi-LSTM: bi-directional long short term memory neural network; RNN: recurrent neural network; IP: industrial production; PLS: partial least squares}

\begin{abstract}
This study leverages narrative from global newspapers to construct theme-based knowledge graphs about world events, demonstrating that features extracted from such graphs improve forecasts of industrial production in three large economies compared to a number of benchmarks. Our analysis relies on a filtering methodology that extracts ``backbones'' of statistically significant edges from large graph data sets. We find that changes in the eigenvector centrality of nodes in such backbones capture shifts in relative importance between different themes significantly better than graph similarity measures. We supplement our results with an interpretability analysis, showing that the theme categories ``disease'' and ``economic'' have the strongest predictive power during the time period that we consider. Our work serves as a blueprint for the construction of parsimonious -- yet informative -- theme-based knowledge graphs to monitor in real time the evolution of relevant phenomena in socio-economic systems.
\end{abstract}


\maketitle
\section{Introduction}

People are natural storytellers who rely on narrative to make decisions, particularly when faced with uncertainty. There is a large body of literature on narrative, with most works spanning disciplines such as psychology, cognitive sciences or political sciences \citep{brosch2013impact, clore2009affective, bruner1990acts, king2017news}. 

Keynes describes the state of mind that results in human actions as ``animal spirits'', which ultimately are reflected in economic indicators such as consumer confidence~\citep{keynes2018general}. Shiller finds that viral narratives spread by newspapers play a causal role in economic activity ~\citep{shiller2017narrative}. Along the same lines, Conviction Narrative Theory (CNT) demonstrates that changes in narrative are a precursor to changes in economic growth~\citep{tuckett2014bringing}. A recent study on CNT suggests that economic agents reassure themselves by building narratives supporting their expectations of the result of their actions~\citep{nyman2018news}.

In recent years, the evolution of big data and natural language processing has enabled the quantification of news narrative and its potential to change the course of social systems. This has allowed to push econometric models such as vector autoregressions~\citep{stock2001vector}, or structural frameworks such as dynamic stochastic general equilibrium models~\citep{christiano2005nominal,smets2007shocks} beyond conventional variables. In this respect, research increasingly explores features derived from narrative and their capacity of improving economic forecasts~\citep{buonoevaluation,elshendy2018using,yang2020knowledge}.

One possible way to capture and quantify narrative is via graphs that map the interactions between concepts or events pertaining to the object of study. A wide range of applications demonstrates that graphs are indeed very successful at capturing complex relationships \citep{emmert2018understanding}. 

This study leverages narrative from global newspapers to construct theme-based knowledge graphs, demonstrating that features derived from such graphs improve forecasts of industrial production (IP) in three large economies. The findings are supported by an interpretability analysis, showing that disease and economy-related themes have the strongest predictive power.

\section{Literature review}
\label{sec: literature review}

This section covers a selection of existing literature on graphs and their applications to the analysis of economic systems. A range of methods has been proposed to extract and interpret the information contained in large graphs, and this section addresses those techniques most pertinent to weighted undirected graphs, which are the object of this study.

Over recent years, there has been increasing interest in examining economic topics in terms of graphs. The study of economic graphs is an interdisciplinary field, spanning areas such as economics, the social sciences, computer science, statistics and business and management~\citep{emmert2018understanding}. Existing studies cover a wide range of economic graphs, each one with their own meaning. For instance, a study on graph centrality and funding rates finds that interbank spreads are significantly affected by measures of centrality, with the effects of graph centrality increasing during the global financial crisis in 2008~\citep{temizsoy2016network}. Constantin \emph{et al}. build a graph of European banks and estimate how negative shocks for one bank's returns depend on the impact of negative shocks on other banks' returns, effectively building an early warning model for bank distress using network effects~\citep{constantin2018network}. A paper by Carvalho \emph{et al}. examines firm-level disturbance after the Japanese tsunami in 2011 and finds that  the propagation of the shock over input-output linkages can account for a 1.2 percentage point decline in Japan’s gross output in the year following the earthquake~\citep{carvalho2016supply}.
Adamic \emph{et al} use established graph analysis tools to describe the time-series dimensions of information and liquidity flows in the E-mini S\&P stock index futures market, showing robust contemporaneous correlations between graph features and financial variables, with the former significantly leading (Granger-causing) intertrade duration, trading volume and other liquidity metrics~\citep{adamic2017trading}. Piccardi and Tajoli show that high centralization observed for complex products determines the hierarchy of the global trade graph as they form an important share of total trade. This implies an uneven distribution of trade links between countries, making them more vulnerable to shocks. The authors conclude that the present structure of the global trade graph is exposed to specific shocks, and quantify the impact of shock propagation from the central nodes~\citep{piccardi2018complexity}.
A study by Bonaccorsi \emph{et al}. measures country centrality in multilayer graphs, demonstrates that these centralities are consistent with a North-South divide and positively correlated with economic variables such as GDP per capita~\citep{bonaccorsi2019country}. Yang \emph{et al}. extract entities from textual data and link them as knowledge graphs to macro variables. The authors show that incorporating the features extracted from knowledge graphs in a macroeconomic forecasting framework significantly improves predictions of macroeconomic variables such as inflation, net export growth or housing prices~\citep{yang2020knowledge}.
Guo and Vargo use themes and location data from the Global Database of Events, Language and Tone (GDELT) to show that population, trade,
cultural proximity, and geographic closeness drive international news attention~\citep{guo2020predictors}.
Stern \emph{et al}. examine intermedia agenda-setting by proposing a method to infer graphs of influence between different news sources on a given subject ~\citep{stern2020network}. The authors find that influence very much depends on the topic, with news outfits being agenda-setters (represented by central nodes) for some topics and  followers (represented by peripheral nodes) for others. Campi \emph{et al}. explore the determinants of specialisation in agricultural production, describing it as a time-sequence of bipartite graphs, linking countries to their produced agricultural products ~\citep{campi2020countries}. The study concludes that agricultural production is a dense graph of well-defined and stable communities of countries and products that are characterised by environmental conditions as well as economic, socio-political and technological factors.
A study by Bellomarini \emph{et al}. analyses the impact of the COVID-19 outbreak on the graph of business relationships between Italian companies and identifies transactions that could lead to the takeover of strategic companies~\citep{bellomarini2020covid}. Gomez \emph{et al}. apply graph analysis to understand business cycle synchronization in the European Union over the last 20 years, observing that co-movements and country interactions have increased notably since the Eurozone crisis in 2011/12, while they remained relatively stable before~\citep{matesanz2017synchronization}.

Graph analysis is an effective means to represent and analyze complex systems.
However, real-world graph data sets can be very large, both in terms of number of nodes and connections between them, making it difficult to identify those elements that are most crucial to the properties of such systems (often referred to as a graph's ``backbone''). The literature proposes a range of approaches for extracting graph backbones. In this respect, Ghalmane \emph{et al}. differentiate between ``coarse-grained'' and filter-based approaches to graph dimensionality reduction~\citep{ghalmane2020extracting}. ``Coarse-grained'' methods are based on the concept of grouping graph nodes according to some criterion and keeping those with the properties of interest, while filter-based approaches define properties for nodes and edges and discard or preserve them based on the statistical significance of these attributes against a null hypothesis for the graph structure. Within the latter category, the filter proposed by Tumminnello \emph{et al}. identifies edges that are statistically significant with respect to a null hypothesis of random interactions in a weighted graph, described by the hypergeometric distribution~\citep{tumminello2011statistically}. 
The ``disparity filter'' proposed by Serrano \emph{et al} works in a similar fashion, and relies on a null hypothesis of nodes distributing their activity uniformly across their neighbours in the graph~\citep{serrano2009extracting}. 
Marcaccioli and Livan recently proposed a filter based on the P\'olya urn model (which includes the disparity filter as a special case), which can be tuned to a given graph's specific topological properties, and demonstrate its effectiveness at preserving statistically significant links in real-world graphs such as the US airport network or the global input-output trade network~\citep{marcaccioli2019polya}.

\subsection{Hypotheses formulation}

In this study, we attempt to forecast industrial production (IP) leveraging data from the Global Database of Events, Language and Tone (GDELT). Specifically, we address two main research questions, i.e., (1) whether socio-economic dynamics can be captured by knowledge graphs based on themes from GDELT, and (2) whether features derived from such graphs are predictive of changes in economic activity. Accordingly, we formulate the two following hypotheses:
\begin{itemize}
    \item $H_1$: Changes in the structural properties of knowledge graphs based on themes from GDELT are reflective of socio-economic changes.
    \item $H_2$: Features derived from GDELT knowledge graphs add value to forecasts of economic activity.
\end{itemize}

By tackling the above questions, we advance the literature on economic graphs in at least three ways. First, we demonstrate how themes from newspaper articles can be incorporated into macroeconomic forecasting models to improve predictions of economic activity. This use of news narrative is -- to the best of our knowledge -- new. Second, we contribute to the literature on economic graphs by proposing a data-driven methodology to operationalise concepts such as information extraction, feature generation and macroeconomic forecasting. We complement our results with an interpretability study showing that disease and economy-related themes have the strongest predictive power. Third, we contribute to research on tracking change in social systems through news narrative by examining the evolution of graph statistics over time.

\section{Data and methods}
This section introduces GDELT as data source and outlines the filtering methodology that is used to isolate relevant signals. The GDELT Project is a research collaboration of Google Ideas, Google Cloud, Google and Google News, the Yahoo! Fellowship at Georgetown University, BBC Monitoring, the National Academies Keck Futures Program, Reed Elsevier's LexisNexis Group, JSTOR, DTIC and the Internet Archive. The project monitors world newspapers from a variety of perspectives, identifying and extracting items such as themes, emotions, locations, organisations and events. GDELT version two incorporates real-time translation from 65 languages and is updated every 15 minutes \citep{leetaru_the_nodate}. It is a public data set available on the Google Cloud Platform.

The GDELT Global Knowledge Graph (GKG) is a software suite that analyses global newspaper articles in real-time to extract entities such as persons, organizations, locations, dates, themes and emotions~\citep{leetaru2014cultural}. The extraction of location data is done through a process called full text geocoding, developed by Leetaru~\citep{leetaru2012fulltext}. This process applies algorithms to parse through a news item and to identify textual mentions of locations using databases of places. Applying the same principle, themes are extracted from news articles using extensive lists of themes. GDELT contains c 13,000 themes from c 47,000 sources, from over a billion news articles scanned since 2015.

For each scanned news article, the themes field contains all themes the GDELT algorithm identifies, represented as a string of labels. GDELT themes are very nuanced and can be linked to distinct categories such as ``economic'', ``disease'' or ``human rights'' (see appendix \ref{sec:gdelt theme categories} for a list of all theme categories). Theme categories describe events or conditions, except for four purely descriptive theme groups (``actor'', ``ethnicity'', ``language'' and ``animal''), which are removed. Table \ref{FIG:1} illustrates the number of themes for the three countries we will consider in our analysis, both in the original data (no. of themes) and the number of themes after removing the descriptive ones (reduced no. of themes).

\begin{center}
\begin{table}[h]
\begin{adjustbox}{width=\columnwidth,center}
\begin{tabular}{|l||*{3}{c|}}\hline
Country & no. of themes & reduced no. of themes\\
\hline
US & 6880 & 3501\\
Germany & 5313 & 2925\\
Japan & 3330 & 1994\\
\hline
\end{tabular}
\caption{\label{tab:no of themes}Number of of GDELT themes}
\end{adjustbox}
\end{table}
\end{center}

\subsection{Predicted variables}
This study models industrial production (IP) for the US, Germany and Japan.
IP is a measure of economic activity, published on a monthly basis. It represents the output of industrial establishments, covers a broad range of sectors and tracks the monthly change in the volume of production output. The countries are selected for representing large, industrialised economies with diversified trade connections in three parts of the world -- America, Europe and Asia.

\subsection{Filtering methodology}
\label{sec:filtering}

We apply the filtering methodology introduced by Tilly \emph{et al}.~\citep{tilly2021forecasting} to extract observations from GDELT's GKG that are pertinent to economic growth. The methodology consists of three steps -- first, a thematic keyword filter, second, a fine-grained filter using a neural network and third, data aggregation.

As a first step, a high level thematic filter based on the keyword ''economic growth'' is applied to GDELT themes to select relevant articles. On inspection of the filtered data by examining 100 randomly chosen original news articles, it becomes clear that this simple keyword filter retains too many news items that are not relevant to ``economic growth''. 

Hence, a further, more precise filter is required as a second step. We use the same bidirectional long short term memory (Bi-LSTM) architecture as proposed in the original filtering methodology. This algorithm is chosen as it exhibits the best performance in terms of precision, recall and $F_1$ score among a range of algorithms explored. For this step, the raw GDELT data is preprocessed. Each string of theme labels is split into lower case tokens. The tokens for every item are then label-encoded so that the themes are given numbers between zero and $N-1$. For out-of-vocabulary words, an ``unknown'' token is assigned. The length for each token sequence is standardized to address the variable length of these sequences by setting a maximum length of 5,000 tokens and padding. Then, 1,000 news items are manually classified into relevant and non-relevant to ``economic growth'' looking up the original text using the url contained in the DocumentIdentifier field (encoded as one or zero, respectively). The encoded themes represent the predictor, and the classification into relevant/non-relevant represent the predicted data, respectively.

Model performance is assessed using $k$-fold cross-validation as it provides a robust estimate of the performance of a model on unseen data. The training data set is divided into ten subsets. Models are trained on all subsets except one which is held out, and performance is evaluated on the held out validation data set. The process is repeated until all subsets have been used as the held out validation set. 

As performance metrics, precision (the number of true positives divided by the number of true positives and false positives), recall (the number of true positives divided by the number of true positives and the number of false negatives) and the $F_1$ score are used:

\[
    F_1 = 2 \times \frac{\mathrm{precision} \times \mathrm{recall}}{\mathrm{precision}+\mathrm{recall}} \ .
\]
Table \ref{tab:performance} shows the performance of the Bi-LSTM classifier.

\begin{center}
\begin{table}[h]
\begin{adjustbox}{width=\columnwidth,center}
\begin{tabular}{|l||*{4}{c|}}\hline

Classifier & Precision & Recall & F1\\
\hline
Bi-LSTM & 0.8853 & 0.9375 &  0.9101\\
\hline
\end{tabular}
\caption{\label{tab:performance}Bi-LSTM performance}
\end{adjustbox}
\end{table}
\end{center}

A long short term memory (LSTM) is a type of recurrent neural network (RNN) structure that can span longer distances without the loss of short term capacity \citep{hochreiter1997long}. The algorithm enforces constant error flow during backpropagation through internal states of special units and addresses the vanishing gradient problem. This issue originates from the repeated use of the recurrent weight in RNNs, and prevents these models from learning long term dependencies.

Like a RNN, a LSTM passes on information as it propagates forward. However, the operations within the LSTM’s cells differ compared to those of a RNN. A LSTM incorporates a cell state that can be considered the ``memory'' as well as a set of gates. These gates control when information enters the memory, when it is output, and when it is forgotten. Through this structure, the LSTM is able to learn what information to keep or to forget during training, and keep relevant information in memory for an extended period.
While a unidirectional LSTM preserves information from one direction as it only runs forward, a Bi-LSTM runs the inputs forwards and backwards simultaneously. Combining two hidden states allows a Bi-LSTM to retain information from both directions at any time \citep{schuster1997bidirectional}.
Research shows that the Bi-LSTM architecture outperforms unidirectional algorithms in tasks where context matters \citep{graves2005framewise}.

Using the filtered data from step two, for each calendar month from March 2015 to December 2020, theme-based undirected weighted graphs are constructed, applying country filters for the US, Germany and Japan, respectively. In these graphs, the themes represent nodes, the co-occurrences of themes represent edges and the count of the co-occurrences represent weights. For comparison purposes, monthly graphs based on unfiltered GDELT data are also generated.

\subsection{Statistical validation of graph elements}

This section outlines the statistical validation procedure we apply to extract the backbone of each monthly graph.

The monthly theme-based graphs are connected and contain c 4,000 nodes and c 1 million edges. Given their large scale and the potentially large amount of noise contained in them, these graphs require a filtering mechanism in order to extract statistically relevant sets of edges, i.e., the so-called ``backbone''. 

To this end, we applied the disparity filter (see Section~\ref{sec: literature review}), the most widely adopted backbone extraction method. The method starts by normalizing the weights of each edge $(i, j)$ with respect to the total strength of node $i$ (i.e., the total weight on edges starting from node $i$), so that they sum up to one. The method's null hypothesis states that the normalised weights belonging to edges of a node $i$ are random and distributed according to the uniform distribution over $[0,1]$. For each edge, the probability $\alpha_{ij}$ of the edge occurring under the null hypothesis, i.e., a $p$-value, is calculated. For edges whose $p$-value is smaller than a set threshold $\alpha$, the null hypothesis is rejected and they are retained in the backbone. In this study, the threshold $\alpha$ is set to 0.05. Since multiple tests are conducted, the resulting $p$-values are adapted according to the Benjamini-Hochberg (BH) procedure to account for multiple hypothesis testing \citep{benjamini2005false}. It should also be noted that, due to the fact that the normalisation of edge weights is done with respect to one of the two nodes they are connected to, each edge needs to be tested twice. The links retained in the backbone are those for which the null hypothesis can be rejected at least once. 

Applying the disparity filter to the theme-based graphs reduces the number of nodes and edges in each monthly graph by c 50\% and c 90\%, respectively, while still returning connected graphs.

\subsection{Graph features}
For each monthly graph, the eigenvector centrality is calculated for all nodes. This measure is appropriate as the graphs are connected and reflects the broader influence of a node on the filtered graph \citep{constantin2018network}.
Then, a \(T \times K\) matrix is constructed, where \(T\) represents the number of observations (i.e. calendar months) and \(K\) stands for the number of nodes represented by themes.

The portrait divergence for each graph compared to the previous month's graph is calculated. This measure is a method for comparing graphs based on the graph's portrait, which encodes the distribution of the shortest-path lengths in a graph. The portrait divergence is a comprehensive measure of how the topological features of two graphs differ \citep{tandardini_2019_comparing}.

\subsection{Explanatory variables}
When forecasting IP, the monthly percentage changes in the baltic dry index and the crude oil price, respectively, are incorporated to control for macroeconomic effects.
The baltic dry index is considered a leading indicator for economic growth, representing global trade volume~\citep{bildirici2015baltic}. Van Eyden \emph{et al}. find that there is a significant link between changes in oil price and economic activity in OECD countries~\citep{van2019oil}. 

\subsection{Data preprocessing}
In this section, we summarize the data preparation methods.

The values for IP are used as predicted variable for each of the three aforementioned economies, with index values reflecting the monthly percentage change. The augmented Dickey Fuller unit root test is applied to 20 years of monthly data for IP as well as the explanatory variables and stationarity is not rejected at 5\% significance for any of them.

Any missing values in the \(T \times K\) eigenvector centralities matrix are filled by zero, which is the lowest possible centrality score. The augmented Dickey Fuller unit root test is applied to eigenvector centralities, portrait divergences and stationary is not rejected at 5\% for any of the variables. The eigenvector centralities are  standardized by removing the mean and scaling to unit variance.

\section{Analysis}
In this section, we set out the analysis conducted to establish if the graph features derived from GDELT data have predictive power.

\subsection{Granger causality analysis}
The Granger causality between the eigenvector centralities and the  predicted variable is assessed to establish if there are statistical relationships between those variables~\citep{granger1969investigating}. Granger causality is testing for precedence rather than true causality, hence it may be found even in the absence of an actual causal connection \citep{leamer_1985}.

For this test, the null hypothesis  stipulates that lagged graph features are not Granger-causing IP at a significance level of 5\%, while the alternate hypothesis states that lagged graph features are Granger-causing IP at the same significance level.

The Granger causality for lags up to a maximum of three months is examined. The $p$-values are adjusted using the Benjamini-Hochberg (BH) procedure to control for multiple hypothesis testing~\citep{benjamini2005false}.

\subsection{Forecasting}
\label{sec:forecasting}
As further step in our analysis, we predict IP for the US, Germany and Japan adopting a factor-augmented autoregressive framework as proposed by Girardi \emph{et al}.~\citep{girardi2016factor}.

As a first step, we predict IP using an autoregressive model including the predicted variable and the explanatory macroeconomic variables. This framework allows modeling a \(T \times K\) multivariate time series \(Y\), where \(T\) denotes the number of observations and \(K\) the number of variables. The framework is defined as
\begin{equation}\label{eq:1}
  Y_{t } =  v + A_{1}Y_{t-1}+\dots +  A_{p}Y_{t-p} + u_{t}
\end{equation}
where \(A_{i}\) is a \(K \times K\) coefficient matrix, $v$ is a constant and $u_{t}$ is white noise.

As it is not possible to incorporate the large number of eigenvector centralities into the autoregressive framework described in Eq.~\eqref{eq:1}, we extract factors from this broad set of features for inclusion into the model. Therefore, as a second step, we apply Partial Least Squares (PLS) to reduce dimensionality and to extract relevant information from the eigenvector centralities. This technique is suitable for data sets whose number of features is considerably larger than the number of observations, and collinearity of features exists~\citep{cubadda2012medium}. PLS incorporates information from predicted variable and predictors when computing scores and loadings, which are chosen to maximise the covariance between predicted variable and predictors~\citep{de1993simpls}.

PLS is implemented on the residuals derived from the autoregressive model including the predicted variable and the explanatory macroeconomic variables only. The residuals include the part of the predicted variable that is not explained and therefore, applying PLS to the eigenvector centralities based on GDELT themes provides additional information to the predictors. The orthogonal relationship between the predicted variable and the residuals maintains the orthogonality between the factors extracted by PLS and the autoregressive components. For each country variable, cross-validation analysis shows that the residual sum of squares is increasing in a model with more than five factors, indicating that five PLS factors are appropriate~\citep{tobias1995introduction}. The first five PLS components account for around 70\% of the variation in the respective predicted variables.

In a third step, for each country, the predicted variable, the explanatory variables and the five PLS components derived from the eigenvector centralities are employed as input into the autoregressive framework described in Eq.~\eqref{eq:1} to form a factor augmented autoregressive model~\citep{colladon2019using}. 

For each model, the the Akaike (AIC) and the Bayesian (BIC) information critera are applied to identify the optimal lag length. Both metrics follow the concept that the inclusion of a further term should improve the model although the model should also be penalised for adding to the number of parameters to be computed. Once the improvement in goodness-of-fit exceeds the penalty term, the statistic related to the information criterion decreases. Therefore, we select the lag that minimises the information criterion~\citep{book}.

We construct three benchmarks for performance comparison -- first, an autoregressive framework including the predicted variable and explanatory macroeconomic variables, second, an autoregressive framework incorporating the predicted variable, explanatory macroeconomic variables and the respective country's portrait divergence scores, and third, an autoregressive framework incorporating the predicted variable, explanatory macroeconomic variables and five PLS components derived from eigenvector centralities based on unfiltered GDELT data.

Performance is assessed using walk-forward cross-validation and the root mean squared error (RMSE). The data set is split into three folds. In the $k\textsuperscript{th}$ split, this cross-validation technique returns the first $k$ folds as training set and the $(k+1)\textsuperscript{th}$ fold as test set, appropriate for time series data.

The modified Diebold Mariano test proposed by Harvey, Leybourne and Newbold~\citep{harvey1997testing} is used to gauge whether model forecasts are significantly different from each other. 

\section{Research findings}
This section presents the findings from the analysis set out in the previous section.

\subsection{Evolution of graph features}
\begin{figure*}
  \includegraphics[width=\textwidth,height=5cm]{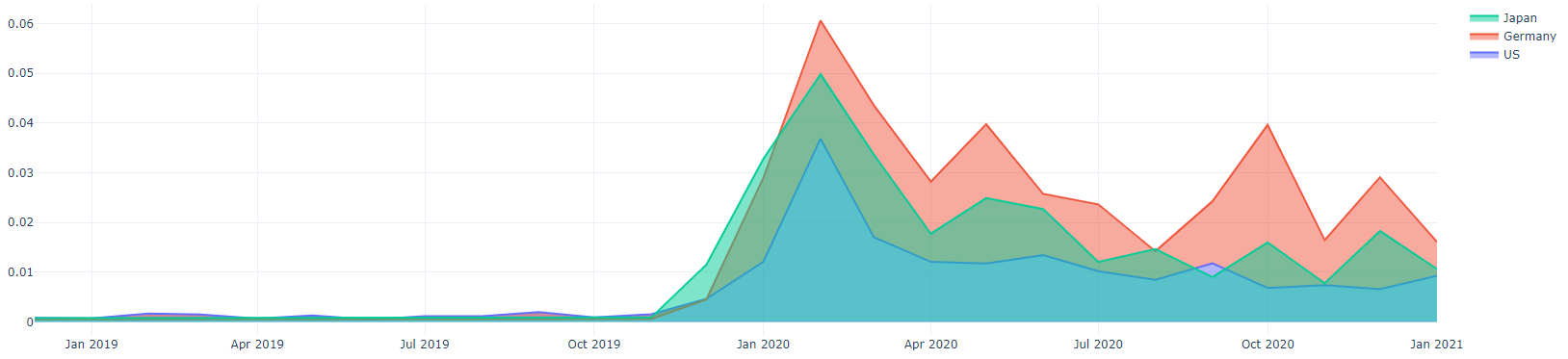}
  \caption{Influence of COVID-19 related themes within monthly graphs.}
  \label{FIG:1}
\end{figure*}

Theme-based knowledge graphs can be employed to identify change in specific aspects of social systems, illustrated here by the the example of the COVID-19 pandemic. Fig \ref{FIG:1} shows the evolution of the median eigenvector centrality of COVID-19 symptom related themes. These themes were selected manually for their relevance, such as ``pneumonia'', ``fever'' or ``cough'', within monthly graphs for the US, Germany and Japan (see section \ref{sec:gdelt covid themes} for a complete list of themes used). The monthly median eigenvector centralities reflect the timing and impact of events in specific geographies on the example of the COVID-19 outbreak in 2020. Prior to end of 2019, the median eigenvector centralities associated with COVID-19 symptoms were consistently low for all three countries, then picking up notably as the virus spread across the globe, first in Japan, then in Germany and at last in the US.\\

\subsection{Granger causality test results}
The eigenvector centralities from theme-based knowledge graphs and the IP index values for three countries are tested for Granger causality, with a maximum lag of three months. Table \ref{tab:ip_granger} exhibits the number of BH-adjusted $p$-values that exhibit significance at 5\% for each country's IP index.

\begin{center}
\begin{table}[h]
\begin{adjustbox}{width=\columnwidth,center}
\begin{tabular}{|l||*{3}{c|}}\hline
\backslashbox{Country}{Data set}
&\makebox[6em]{GC (filt)}&\makebox[6em]{Reverse GC}&\makebox[6em]{GC (unfilt)}\\\hline
US & 673  & 583 & 94\\  
Germany & 783  & 742 & 67\\
Japan & 552 & 556 & 99\\
\hline
\end{tabular}
\caption{IP: Number of significant BH-adjusted $p$-values. GC refers to Granger causality; filt (unfilt) refers to filtered (unfiltered) GDELT data}
\label{tab:ip_granger}
\end{adjustbox}
\end{table}
\end{center}

The Granger causality analysis results in a considerably larger number of statistically significant BH-adjusted $p$-values for those features generated from filtered GDELT data compared to those from unfiltered GDELT data, suggesting that the filtering methodology applied in section \ref{sec:filtering} isolates relevant signals.

Strong reverse Granger causality exists between IP and the eigenvector centralities generated from monthly theme-based knowledge graphs. This is unsurprising given that press reporting is bi-directional. Events -- extracted by GDELT as themes -- covered by global newspapers impact macroeconomic variables such as IP while the state of the economy -- as measured in macroeconomic variables -- is commented on in the press.

\subsection{Forecast error analysis}
\label{sec:forecast error analysis}
For the US, Germany and Japan, the respective eigenvector centrality matrices are condensed into five factors applying PLS. They are then incorporated into the forecasting framework described in section \ref{sec:forecasting} to predict IP. All models have a lag of one month, determined by evaluating AIC and BIC.

The columns of Table \ref{tab:IP_results} show the performance metric (RMSE) from the factor augmented models compared to three benchmarks, which are based on autoregressive models -- the first contains the predicted variable and explanatory macroeconomic variables only (referred to as BM1), the second includes the predicted variable, explanatory macroeconomic variables and portrait divergence scores (referred to as BM2) and the third incorporates the predicted variable, explanatory macroeconomic variables and five components from eigenvector centralities derived from unfiltered GDELT data (BM3).  The numbers in the cells represent the RMSE in percentage terms for each model and its benchmarks. Blue (red) cells denote cases in which the models outperform (underperform) the respective benchmarks. In the column ``Sign.'', numbers in parentheses correspond to the number of significant coefficients associated with GDELT factors in the model in Eq. (\ref{eq:1}), with the asterisks denoting the level of their statistical significance.

\begin{center}
\begin{table}[h]
\begin{adjustbox}{width=\columnwidth,center}
\begin{tabular}{|l||*{6}{c|}}\hline
\backslashbox{IP for}
&\makebox[4em]{Model}&\makebox[4em]{BM1}& \makebox[4em]{BM2}& \makebox[4em]{BM3}&\makebox[4em]{Sign.}\\\hline

US & 1.6139 & \cellcolor{blue!15} 1.6341& \cellcolor{blue!15}1.6376& \cellcolor{blue!15}1.6144&***(1), **(1)\\

Germany & 2.6942 & \cellcolor{blue!15}2.7079 & \cellcolor{blue!15}2.7082 & \cellcolor{blue!15}2.7159 & *(1)\\

Japan & 2.4978 & \cellcolor{blue!15}2.5300 & \cellcolor{blue!15}2.5321& \cellcolor{blue!15}2.5049&*(1)\\
\hline
\end{tabular}
\caption{Results of the model in Eq. \ref{eq:1} applied to IP. Numbers represent the RMSE (\%). Blue (red) cells denote cases in which the model outperforms (underperforms) the benchmark. Numbers in parentheses correspond to the number of significant coefficients associated with GDELT factors in the model in Eq. (\ref{eq:1}) (${***}$ denotes at least one GDELT sentiment factor with $p$-value $<$ 0.01, ${**}$ $<$ 0.05, ${*}$ $<$ 0.1).}
\label{tab:IP_results}
\end{adjustbox}
\end{table}
\end{center}

Rows one to three in Table \ref{tab:IP_results} show that the models containing the filtered GDELT factors outperform all three benchmarks for the US, Germany and Japan, respectively. All factor augmented models contain at least one statistically significant GDELT factor at 0.1, 0.05 and 0.01, each. 

In 2020, IP for all three countries experienced high levels of volatility as the lockdowns imposed by governments around the world severely curbed economic activity. In the cross-validation, the last validation set incorporates the period of the COVID-19 outbreak in 2020. Predictions on this last validation set exhibit a much larger error metric across countries than those predictions on the validation sets that exclude the outbreak. However, performance dynamics during the COVID-19 outbreak remain the same in that the factor enhanced models outperform their benchmarks.

Table \ref{tab:DM_IP_results} contains the $p$-values from the modified Diebold Mariano test. This test is used to to gauge if factor enhanced model forecasts are significantly different from the benchmark predictions as set out in section \ref{sec:forecast error analysis}.

\begin{center}
\begin{table}[H]
\begin{adjustbox}{width=\columnwidth,center}
\begin{tabular}{|l||*{5}{c|}}\hline
\backslashbox{IP for}{Data set}
&\makebox[6em]{Model - BM1}& \makebox[6em]{Model - BM2}& \makebox[6em]{Model - BM3}\\\hline
US & 0.0000 & 0.0000 & 0.0000\\
Germany & 0.0103 & 0.0093 & 0.0001\\
Japan & 0.0000 & 0.0000 & 0.0887\\
\hline

\hline
\end{tabular}
\caption{$p$-values from modified Diebold Mariano test}
\label{tab:DM_IP_results}
\end{adjustbox}
\end{table}
\end{center}

According to the modified Diebold Mariano test, all model forecasts for IP are statistically different to BM1, BM2 and BM3 predictions at 1 \% or 10 \% significance, respectively.

Results suggest that features derived from narrative-based knowledge graphs improve IP forecasts for three large economies. In particular, the factor augmented autoregressive models deliver consistently better performance compared to those models that contain a single graph similarity measure (BM2) or factors based on unfiltered GDELT data (BM3). This indicates that eigenvector centralities are sufficiently nuanced and thus well suited to capturing the changing dynamics in theme-based graphs while this is not achieved by the portrait divergence, which is a single score measuring the change in topological properties.

\subsection{Drivers of GDELT factors}
In this section, we examine the loadings corresponding to each component to gain an understanding of the relationship between themes and the PLS components extracted from GDELT data. Loadings correspond to the strength of relationship between the original eigenvector centralities for each graph node (represented by a GDELT theme) and the PLS components, quantifying the importance of the underlying themes in each of them.

%
%
%

\begin{figure}[pos=h!]
	\centering
	\includegraphics[scale=.6]{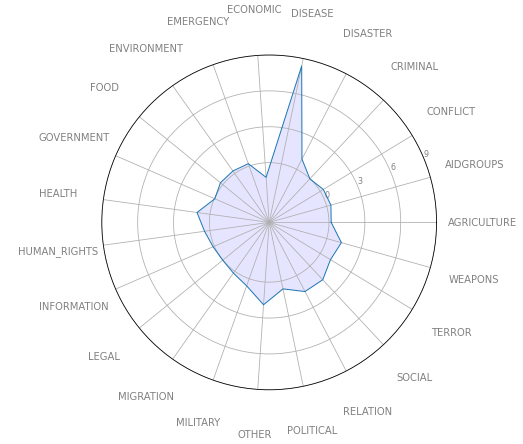}
	\includegraphics[scale=.6]{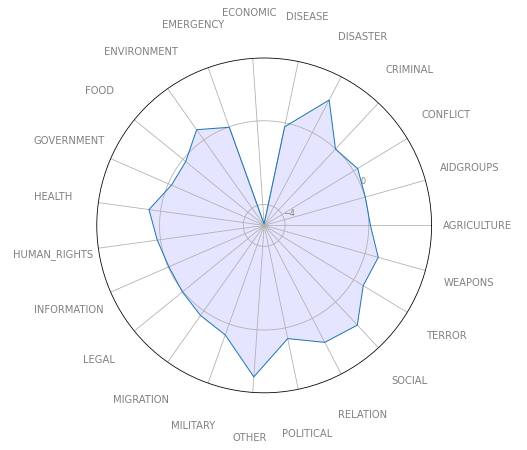}
	\caption{Top: Significant PLS components explained by theme categories (Factor 1, US). Bottom: Significant PLS components explained by theme categories (Factor 4, US)}
	\label{FIG:2}
\end{figure}

GDELT themes represent specific events or conditions such as `'inflation'', ``economic growth'' or ``disease'' and are mapped to 22 distinct theme categories using the GDELT naming convention. For example, themes containing the word ``disease'' are mapped to the disease category; themes including the word ``weapon'' are mapped to the weapons category, etc (see section \ref{sec:gdelt theme categories} for detailed list).

For each component, the loadings are summed according to the 22 distinct theme categories set out above. Mapping GDELT themes to theme categories delivers insights into the relationship of these categories with each PLS component.

In Fig. \ref{FIG:2} we show radar charts of the theme categories associated with the loadings of the statistically significant PLS components used to forecast IP in the US. Table \ref{tab:IP_results} illustrates that the corresponding model outperforms the three benchmarks and exhibits substantial statistical significance.

This example demonstrates that the factors we use to model IP can be linked to distinct theme categories. Thus, change in such categories over time helps explain movements in IP, which is a monthly proxy of economic activity.

Of the 22 distinct theme categories, ``disease'' and ``economic'' exhibit the strongest relationships for factor 1 and factor 4, respectively and therefore are their most important drivers with the strongest predictive power.
This also applies to the statistically significant PLS components for the models predicting German and Japanese IP, which are driven by ``economic'' and ``disease'' categories, respectively. These charts can be made available on request.

\section{Discussion}
Our study presents a novel method of incorporating themes from global newspaper narrative into macroeconomic forecasts.
We apply an effective filtering methodology for extracting relevant signals from a large volume of data.
The filtered data is used to build theme-based weighted undirected knowledge graphs for each calendar month and for three large economies, respectively. For comparison purposes, monthly graphs based on unfiltered GDELT data are generated.
For each monthly graph, we calculate the eigenvector centralities and portrait divergences, creating a monthly frequency time series of features.
On the example of the COVID-19 outbreak end of 2019, we show that the change in graph centralities is indicative of genuine real-world change. This finding supports hypothesis $H_1$.

The monthly eigenvector centralities for each of the three economies exhibit robust Granger causality, indicating a statistical relationship between graph features and the IP values in corresponding countries. For each country, we reduce the eigenvector centralities into five components applying PLS and incorporate them into a factor augmented autoregressive framework. Results show that features derived from theme-based graphs significantly improve IP forecasts for the US, Germany and Japan. In particular, the factor augmented models consistently outperform the benchmark models that contain a single graph similarity measure only (portrait divergence) as well as the benchmark models that incorporate five PLS components derived from unfiltered GDELT data. 
These results support hypothesis $H_2$.
Grouping over one thousand GDELT themes into 22 distinct theme categories helps interpret the relationship between those categories and each PLS factor, with themes related to ``disease'' and ``economic'' being their most important drivers.

Our work makes contributions to three main streams of research. First, our study expands existing research on using big data and machine learning in macroeconomics \citep{giannone2008nowcasting, mccracken2016fred, baker2016measuring, coulombe2020machine}. We leverage machine learning, graph analytics and natural language processing to represent the state of the economy in terms of themes derived from news narrative, and demonstrate how these themes can be incorporated into macroeconomic forecasting models to improve predictions of economic activity. This use of news narrative is -- to the best of our knowledge -- new. Second, we contribute to the literature on economic graphs, which have a wide range of applications and often span across several disciplines~ \citep{emmert2018understanding}. Our work adds to existing research on information extraction, feature generation and macroeconomic forecasting using textual data at scale, presenting a data-driven approach to operationalising such concepts and to integrating them into macroeconomic forecasting frameworks. We supplement our findings with an interpretability analysis illustrating that disease and economy-related themes have the strongest predictive power. Third, we expand the research on tracking change in social systems through narrative. As nodes in a graph, themes convey detailed information about the state of a system, while the evolution of their properties over time is indicative of changes. These narrative-based graphs can be modified to track specific phenomena such as ``inflation'', ``migration'' or ``human rights violations'' in real-time and, therefore, have a broad range of applications. 

\subsection{Limitations and ideas for further research}
Some potential limitations of our study should be acknowledged. First, the forecasting framework focuses on linear relationships between predictors and predicted variables. Exploring non-linear modelling techniques may generate further understanding of the interaction between these variables and could be an extension to this project.
Second, our work represents a proof of concept and does not focus on performance optimisation. In addition, our study is limited to predicting one macroeconomic variable for three large economies. Its pertinence to real-world applications could be enhanced by expanding to a broader range of variables and economies. 
Third, GDELT has a relatively short track record, going back to February 2015. The limited amount of observations is likely to affect results, in particular when modelling monthly frequency data. 
Fourth, this study explores a limited number of graph features and graph analysis techniques. Our work could be expanded by considering alternative centrality and similarity measures for inclusion into the forecasting framework.

\section{Conclusions}
This study proposes a new method of incorporating themes derived from global newspaper narrative into macroeconomic forecasts, and demonstrates that factors derived from these themes significantly improve predictions of economic activity for three large economies.
The interpretation of the factors extracted from eigenvector centralities shows that themes associated with the ``disease'' and ``economic'' categories have the strongest predictive power and therefore help explain changes in economic activity.
The theme-based knowledge graphs used in this study represent a nuanced picture of the state of the economy at a given point in time. They can be modified to monitor specific aspects of social systems in real-time and thus have a wide range of applications.

\section{Appendix}

\subsection{COVID-19 related themes}
\label{sec:gdelt covid themes}
The themes listed below represent COVID-19 related symptoms. Applying these themes, we calculate the median eigenvector centralities for each calendar month and illustrate the evolution of the eigenvector centralities over time on the example of the COVID-19 outbreak.

\begin{itemize}
    \item \texttt{WB\_2165\_HEALTH\_EMERGENCIES}
    \item \texttt{WB\_1406\_DISEASES}  
    \item \texttt{TAX\_DISEASE\_CORONAVIRUS}
    \item \texttt{TAX\_DISEASE\_EPIDEMIC}
    \item \texttt{TAX\_DISEASE\_OUTBREAK}
    \item \texttt{TAX\_DISEASE\_INFECTION}
    \item \texttt{TAX\_DISEASE\_PNEUMONIA}
    \item \texttt{TAX\_DISEASE\_FEVER}
    \item \texttt{TAX\_DISEASE\_INFECTIOUS}
    \item \texttt{TAX\_DISEASE\_FLU}
    \item \texttt{TAX\_DISEASE\_COUGH}

\end{itemize}

\subsection{GDELT theme categories}
\label{sec:gdelt theme categories}
All GDELT themes are assigned to one of 26 theme categories. For example, the ``Weapons'' category includes 81 themes such as \texttt{``TAX\_WEAPONS\_GUNS''}, \texttt{``TAX\_WEAPONS\_BOMB''} and\\ \texttt{``TAX\_WEAPONS\_SUICIDE\_BOMB''}. We removed themes belonging to ``Actor'', ``Language'', ``Animal'' and ``Ethnicity'' for our analysis as they are purely descriptive.

\begin{itemize}
  \item Economic
  \item Disease
  \item Actor
  \item Language
  \item Ethnicity
  \item Animal
  \item Disaster
  \item Social
  \item Relation
  \item Political
  \item Health
  \item Weapons
  \item Military
  \item Terror
  \item Environment
  \item Food
  \item Government
  \item Aid groups
  \item Information
  \item Conflict
  \item Emergency
  \item Human rights
  \item Migration
  \item Legal
  \item Criminal
  \item Other (various events or conditions, c 6\% of themes)\\
\end{itemize}

\section*{Declaration of competing interest}
The authors declare that they have no known competing financial interests or personal relationships that could have appeared to influence the work reported in this paper.

\section*{Acknowledgments}
The authors thank Fabio Caccioli and David Tuckett for very helpful feedback on preliminary versions of our manuscript.

\bibliographystyle{apacite}

\bibliography{cas-refs}

\begin{thebibliography}{}

\bibitem [\protect \citeauthoryear {%
Adamic%
, Brunetti%
, Harris%
\BCBL {}\ \BBA {} Kirilenko%
}{%
Adamic%
\ \protect \BOthers {.}}{%
{\protect \APACyear {2017}}%
}]{%
adamic2017trading}
\APACinsertmetastar {%
adamic2017trading}%
\begin{APACrefauthors}%
Adamic, L.%
, Brunetti, C.%
, Harris, J\BPBI H.%
\BCBL {}\ \BBA {} Kirilenko, A.%
\end{APACrefauthors}%
\unskip\
\newblock
\APACrefYearMonthDay{2017}{}{}.
\newblock
{\BBOQ}\APACrefatitle {Trading networks} {Trading networks}.{\BBCQ}
\newblock
\APACjournalVolNumPages{The Econometrics Journal}{20}{3}{S126--S149}.
\newblock
\begin{APACrefDOI} \doi{https://doi.org/10.1111/ectj.12090} \end{APACrefDOI}
\PrintBackRefs{\CurrentBib}

\bibitem [\protect \citeauthoryear {%
Baker%
, Bloom%
\BCBL {}\ \BBA {} Davis%
}{%
Baker%
\ \protect \BOthers {.}}{%
{\protect \APACyear {2016}}%
}]{%
baker2016measuring}
\APACinsertmetastar {%
baker2016measuring}%
\begin{APACrefauthors}%
Baker, S.%
, Bloom, N.%
\BCBL {}\ \BBA {} Davis, S\BPBI J.%
\end{APACrefauthors}%
\unskip\
\newblock
\APACrefYearMonthDay{2016}{}{}.
\newblock
{\BBOQ}\APACrefatitle {Measuring economic policy uncertainty} {Measuring
  economic policy uncertainty}.{\BBCQ}
\newblock
\APACjournalVolNumPages{The quarterly journal of
  economics}{131}{4}{1593--1636}.
\newblock
\begin{APACrefDOI} \doi{https://doi.org/10.1093/qje/qjw024} \end{APACrefDOI}
\PrintBackRefs{\CurrentBib}

\bibitem [\protect \citeauthoryear {%
Bellomarini%
\ \protect \BOthers {.}}{%
Bellomarini%
\ \protect \BOthers {.}}{%
{\protect \APACyear {2020}}%
}]{%
bellomarini2020covid}
\APACinsertmetastar {%
bellomarini2020covid}%
\begin{APACrefauthors}%
Bellomarini, L.%
, Benedetti, M.%
, Gentili, A.%
, Laurendi, R.%
, Magnanimi, D.%
, Muci, A.%
\BCBL {}\ \BBA {} Sallinger, E.%
\end{APACrefauthors}%
\unskip\
\newblock
\APACrefYearMonthDay{2020}{}{}.
\newblock
{\BBOQ}\APACrefatitle {COVID-19 and company knowledge graphs: assessing golden
  powers and economic impact of selective lockdown via AI reasoning} {Covid-19
  and company knowledge graphs: assessing golden powers and economic impact of
  selective lockdown via ai reasoning}.{\BBCQ}
\newblock
\APACjournalVolNumPages{arXiv preprint arXiv:2004.10119}{}{}{}.
\PrintBackRefs{\CurrentBib}

\bibitem [\protect \citeauthoryear {%
Benjamini%
\ \BBA {} Yekutieli%
}{%
Benjamini%
\ \BBA {} Yekutieli%
}{%
{\protect \APACyear {2005}}%
}]{%
benjamini2005false}
\APACinsertmetastar {%
benjamini2005false}%
\begin{APACrefauthors}%
Benjamini, Y.%
\BCBT {}\ \BBA {} Yekutieli, D.%
\end{APACrefauthors}%
\unskip\
\newblock
\APACrefYearMonthDay{2005}{}{}.
\newblock
{\BBOQ}\APACrefatitle {False discovery rate--adjusted multiple confidence
  intervals for selected parameters} {False discovery rate--adjusted multiple
  confidence intervals for selected parameters}.{\BBCQ}
\newblock
\APACjournalVolNumPages{Journal of the American Statistical
  Association}{100}{469}{71--81}.
\newblock
\begin{APACrefDOI} \doi{https://doi.org/10.1198/016214504000001907}
  \end{APACrefDOI}
\PrintBackRefs{\CurrentBib}

\bibitem [\protect \citeauthoryear {%
Bildirici%
, Kay{\i}k{\c{c}}{\i}%
\BCBL {}\ \BBA {} Onat%
}{%
Bildirici%
\ \protect \BOthers {.}}{%
{\protect \APACyear {2015}}%
}]{%
bildirici2015baltic}
\APACinsertmetastar {%
bildirici2015baltic}%
\begin{APACrefauthors}%
Bildirici, M\BPBI E.%
, Kay{\i}k{\c{c}}{\i}, F.%
\BCBL {}\ \BBA {} Onat, I\BPBI {\c{S}}.%
\end{APACrefauthors}%
\unskip\
\newblock
\APACrefYearMonthDay{2015}{}{}.
\newblock
{\BBOQ}\APACrefatitle {Baltic Dry Index as a major economic policy indicator:
  the relationship with economic growth} {Baltic dry index as a major economic
  policy indicator: the relationship with economic growth}.{\BBCQ}
\newblock
\APACjournalVolNumPages{Procedia-Social and Behavioral
  Sciences}{210}{}{416--424}.
\newblock
\begin{APACrefDOI} \doi{https://doi.org/10.1016/j.sbspro.2015.11.389}
  \end{APACrefDOI}
\PrintBackRefs{\CurrentBib}

\bibitem [\protect \citeauthoryear {%
Bonaccorsi%
, Riccaboni%
, Fagiolo%
\BCBL {}\ \BBA {} Santoni%
}{%
Bonaccorsi%
\ \protect \BOthers {.}}{%
{\protect \APACyear {2019}}%
}]{%
bonaccorsi2019country}
\APACinsertmetastar {%
bonaccorsi2019country}%
\begin{APACrefauthors}%
Bonaccorsi, G.%
, Riccaboni, M.%
, Fagiolo, G.%
\BCBL {}\ \BBA {} Santoni, G.%
\end{APACrefauthors}%
\unskip\
\newblock
\APACrefYearMonthDay{2019}{}{}.
\newblock
{\BBOQ}\APACrefatitle {Country centrality in the international multiplex
  network} {Country centrality in the international multiplex network}.{\BBCQ}
\newblock
\APACjournalVolNumPages{Applied Network Science}{4}{1}{1--42}.
\newblock
\begin{APACrefDOI} \doi{https://doi.org/10.1007/s41109-019-0207-3}
  \end{APACrefDOI}
\PrintBackRefs{\CurrentBib}

\bibitem [\protect \citeauthoryear {%
Brooks%
\ \BBA {} Tsolacos%
}{%
Brooks%
\ \BBA {} Tsolacos%
}{%
{\protect \APACyear {2010}}%
}]{%
book}
\APACinsertmetastar {%
book}%
\begin{APACrefauthors}%
Brooks, C.%
\BCBT {}\ \BBA {} Tsolacos, S.%
\end{APACrefauthors}%
\unskip\
\newblock
\APACrefYear{2010}.
\newblock
\APACrefbtitle {Real Estate Modelling and Forecasting} {Real estate modelling
  and forecasting}.
\newblock
\begin{APACrefDOI} \doi{https://doi.org/10.1017/CBO9780511814235}
  \end{APACrefDOI}
\PrintBackRefs{\CurrentBib}

\bibitem [\protect \citeauthoryear {%
Brosch%
, Scherer%
, Grandjean%
\BCBL {}\ \BBA {} Sander%
}{%
Brosch%
\ \protect \BOthers {.}}{%
{\protect \APACyear {2013}}%
}]{%
brosch2013impact}
\APACinsertmetastar {%
brosch2013impact}%
\begin{APACrefauthors}%
Brosch, T.%
, Scherer, K\BPBI R.%
, Grandjean, D\BPBI M.%
\BCBL {}\ \BBA {} Sander, D.%
\end{APACrefauthors}%
\unskip\
\newblock
\APACrefYearMonthDay{2013}{}{}.
\newblock
{\BBOQ}\APACrefatitle {The impact of emotion on perception, attention, memory,
  and decision-making} {The impact of emotion on perception, attention, memory,
  and decision-making}.{\BBCQ}
\newblock
\APACjournalVolNumPages{Swiss medical weekly}{143}{}{w13786}.
\newblock
\begin{APACrefDOI} \doi{https://doi.org/10.4414/smw.2013.13786}
  \end{APACrefDOI}
\PrintBackRefs{\CurrentBib}

\bibitem [\protect \citeauthoryear {%
Bruner%
}{%
Bruner%
}{%
{\protect \APACyear {1990}}%
}]{%
bruner1990acts}
\APACinsertmetastar {%
bruner1990acts}%
\begin{APACrefauthors}%
Bruner, J\BPBI S.%
\end{APACrefauthors}%
\unskip\
\newblock
\APACrefYear{1990}.
\newblock
\APACrefbtitle {Acts of meaning} {Acts of meaning}\ (\BVOL~3).
\newblock
\APACaddressPublisher{}{Harvard University Press}.
\PrintBackRefs{\CurrentBib}

\bibitem [\protect \citeauthoryear {%
Buono%
, Kapetanios%
, Marcellino%
, Mazzi%
\BCBL {}\ \BBA {} Papailias%
}{%
Buono%
\ \protect \BOthers {.}}{%
{\protect \APACyear {2018}}%
}]{%
buonoevaluation}
\APACinsertmetastar {%
buonoevaluation}%
\begin{APACrefauthors}%
Buono, D.%
, Kapetanios, G.%
, Marcellino, M.%
, Mazzi, G\BPBI L.%
\BCBL {}\ \BBA {} Papailias, F.%
\end{APACrefauthors}%
\unskip\
\newblock
\APACrefYearMonthDay{2018}{}{}.
\newblock
{\BBOQ}\APACrefatitle {Evaluation of Nowcasting/Flash Estimation based on a Big
  Set of Indicators} {Evaluation of nowcasting/flash estimation based on a big
  set of indicators}.{\BBCQ}.
\PrintBackRefs{\CurrentBib}

\bibitem [\protect \citeauthoryear {%
Campi%
, Due{\~n}as%
\BCBL {}\ \BBA {} Fagiolo%
}{%
Campi%
\ \protect \BOthers {.}}{%
{\protect \APACyear {2020}}%
}]{%
campi2020countries}
\APACinsertmetastar {%
campi2020countries}%
\begin{APACrefauthors}%
Campi, M.%
, Due{\~n}as, M.%
\BCBL {}\ \BBA {} Fagiolo, G.%
\end{APACrefauthors}%
\unskip\
\newblock
\APACrefYearMonthDay{2020}{}{}.
\newblock
{\BBOQ}\APACrefatitle {How do countries specialize in agricultural production?
  A complex network analysis of the global agricultural product space} {How do
  countries specialize in agricultural production? a complex network analysis
  of the global agricultural product space}.{\BBCQ}
\newblock
\APACjournalVolNumPages{Environmental Research Letters}{15}{12}{124006}.
\newblock
\begin{APACrefDOI} \doi{https://doi.org/10.1088/1748-9326/abc2f6}
  \end{APACrefDOI}
\PrintBackRefs{\CurrentBib}

\bibitem [\protect \citeauthoryear {%
Carvalho%
, Nirei%
, Saito%
\BCBL {}\ \BBA {} Tahbaz-Salehi%
}{%
Carvalho%
\ \protect \BOthers {.}}{%
{\protect \APACyear {2016}}%
}]{%
carvalho2016supply}
\APACinsertmetastar {%
carvalho2016supply}%
\begin{APACrefauthors}%
Carvalho, V\BPBI M.%
, Nirei, M.%
, Saito, Y.%
\BCBL {}\ \BBA {} Tahbaz-Salehi, A.%
\end{APACrefauthors}%
\unskip\
\newblock
\APACrefYearMonthDay{2016}{}{}.
\newblock
{\BBOQ}\APACrefatitle {Supply chain disruptions: Evidence from the great east
  japan earthquake} {Supply chain disruptions: Evidence from the great east
  japan earthquake}.{\BBCQ}
\newblock
\APACjournalVolNumPages{Columbia Business School Research Paper}{}{17-5}{}.
\newblock
\begin{APACrefDOI} \doi{http://dx.doi.org/10.2139/ssrn.2883800}
  \end{APACrefDOI}
\PrintBackRefs{\CurrentBib}

\bibitem [\protect \citeauthoryear {%
Christiano%
, Eichenbaum%
\BCBL {}\ \BBA {} Evans%
}{%
Christiano%
\ \protect \BOthers {.}}{%
{\protect \APACyear {2005}}%
}]{%
christiano2005nominal}
\APACinsertmetastar {%
christiano2005nominal}%
\begin{APACrefauthors}%
Christiano, L\BPBI J.%
, Eichenbaum, M.%
\BCBL {}\ \BBA {} Evans, C\BPBI L.%
\end{APACrefauthors}%
\unskip\
\newblock
\APACrefYearMonthDay{2005}{}{}.
\newblock
{\BBOQ}\APACrefatitle {Nominal rigidities and the dynamic effects of a shock to
  monetary policy} {Nominal rigidities and the dynamic effects of a shock to
  monetary policy}.{\BBCQ}
\newblock
\APACjournalVolNumPages{Journal of political Economy}{113}{1}{1--45}.
\newblock
\begin{APACrefDOI} \doi{https://doi.org/10.1086/426038} \end{APACrefDOI}
\PrintBackRefs{\CurrentBib}

\bibitem [\protect \citeauthoryear {%
Clore%
\ \BBA {} Palmer%
}{%
Clore%
\ \BBA {} Palmer%
}{%
{\protect \APACyear {2009}}%
}]{%
clore2009affective}
\APACinsertmetastar {%
clore2009affective}%
\begin{APACrefauthors}%
Clore, G\BPBI L.%
\BCBT {}\ \BBA {} Palmer, J.%
\end{APACrefauthors}%
\unskip\
\newblock
\APACrefYearMonthDay{2009}{}{}.
\newblock
{\BBOQ}\APACrefatitle {Affective guidance of intelligent agents: How emotion
  controls cognition} {Affective guidance of intelligent agents: How emotion
  controls cognition}.{\BBCQ}
\newblock
\APACjournalVolNumPages{Cognitive systems research}{10}{1}{21--30}.
\newblock
\begin{APACrefDOI} \doi{https://doi.org/10.1016/j.cogsys.2008.03.002}
  \end{APACrefDOI}
\PrintBackRefs{\CurrentBib}

\bibitem [\protect \citeauthoryear {%
Colladon%
, Guardabascio%
\BCBL {}\ \BBA {} Innarella%
}{%
Colladon%
\ \protect \BOthers {.}}{%
{\protect \APACyear {2019}}%
}]{%
colladon2019using}
\APACinsertmetastar {%
colladon2019using}%
\begin{APACrefauthors}%
Colladon, A\BPBI F.%
, Guardabascio, B.%
\BCBL {}\ \BBA {} Innarella, R.%
\end{APACrefauthors}%
\unskip\
\newblock
\APACrefYearMonthDay{2019}{}{}.
\newblock
{\BBOQ}\APACrefatitle {Using social network and semantic analysis to analyze
  online travel forums and forecast tourism demand} {Using social network and
  semantic analysis to analyze online travel forums and forecast tourism
  demand}.{\BBCQ}
\newblock
\APACjournalVolNumPages{Decision Support Systems}{123}{}{113075}.
\newblock
\begin{APACrefDOI} \doi{https://doi.org/10.1016/j.dss.2019.113075}
  \end{APACrefDOI}
\PrintBackRefs{\CurrentBib}

\bibitem [\protect \citeauthoryear {%
Constantin%
, Peltonen%
\BCBL {}\ \BBA {} Sarlin%
}{%
Constantin%
\ \protect \BOthers {.}}{%
{\protect \APACyear {2018}}%
}]{%
constantin2018network}
\APACinsertmetastar {%
constantin2018network}%
\begin{APACrefauthors}%
Constantin, A.%
, Peltonen, T\BPBI A.%
\BCBL {}\ \BBA {} Sarlin, P.%
\end{APACrefauthors}%
\unskip\
\newblock
\APACrefYearMonthDay{2018}{}{}.
\newblock
{\BBOQ}\APACrefatitle {Network linkages to predict bank distress} {Network
  linkages to predict bank distress}.{\BBCQ}
\newblock
\APACjournalVolNumPages{Journal of Financial Stability}{35}{}{226--241}.
\newblock
\begin{APACrefDOI} \doi{https://doi.org/10.1016/j.jfs.2016.10.011}
  \end{APACrefDOI}
\PrintBackRefs{\CurrentBib}

\bibitem [\protect \citeauthoryear {%
Coulombe%
, Leroux%
, Stevanovic%
\BCBL {}\ \BBA {} Surprenant%
}{%
Coulombe%
\ \protect \BOthers {.}}{%
{\protect \APACyear {2020}}%
}]{%
coulombe2020machine}
\APACinsertmetastar {%
coulombe2020machine}%
\begin{APACrefauthors}%
Coulombe, P\BPBI G.%
, Leroux, M.%
, Stevanovic, D.%
\BCBL {}\ \BBA {} Surprenant, S.%
\end{APACrefauthors}%
\unskip\
\newblock
\APACrefYearMonthDay{2020}{}{}.
\newblock
{\BBOQ}\APACrefatitle {How is machine learning useful for macroeconomic
  forecasting?} {How is machine learning useful for macroeconomic
  forecasting?}{\BBCQ}
\newblock
\APACjournalVolNumPages{arXiv preprint arXiv:2008.12477}{}{}{}.
\PrintBackRefs{\CurrentBib}

\bibitem [\protect \citeauthoryear {%
Cubadda%
\ \BBA {} Guardabascio%
}{%
Cubadda%
\ \BBA {} Guardabascio%
}{%
{\protect \APACyear {2012}}%
}]{%
cubadda2012medium}
\APACinsertmetastar {%
cubadda2012medium}%
\begin{APACrefauthors}%
Cubadda, G.%
\BCBT {}\ \BBA {} Guardabascio, B.%
\end{APACrefauthors}%
\unskip\
\newblock
\APACrefYearMonthDay{2012}{}{}.
\newblock
{\BBOQ}\APACrefatitle {A medium-N approach to macroeconomic forecasting} {A
  medium-n approach to macroeconomic forecasting}.{\BBCQ}
\newblock
\APACjournalVolNumPages{Economic Modelling}{29}{4}{1099--1105}.
\newblock
\begin{APACrefDOI} \doi{https://doi.org/10.1016/j.econmod.2012.03.027}
  \end{APACrefDOI}
\PrintBackRefs{\CurrentBib}

\bibitem [\protect \citeauthoryear {%
De~Jong%
}{%
De~Jong%
}{%
{\protect \APACyear {1993}}%
}]{%
de1993simpls}
\APACinsertmetastar {%
de1993simpls}%
\begin{APACrefauthors}%
De~Jong, S.%
\end{APACrefauthors}%
\unskip\
\newblock
\APACrefYearMonthDay{1993}{}{}.
\newblock
{\BBOQ}\APACrefatitle {SIMPLS: an alternative approach to partial least squares
  regression} {Simpls: an alternative approach to partial least squares
  regression}.{\BBCQ}
\newblock
\APACjournalVolNumPages{Chemometrics and intelligent laboratory
  systems}{18}{3}{251--263}.
\newblock
\begin{APACrefDOI} \doi{https://doi.org/10.1016/0169-7439(93)85002-X}
  \end{APACrefDOI}
\PrintBackRefs{\CurrentBib}

\bibitem [\protect \citeauthoryear {%
Elshendy%
, Colladon%
, Battistoni%
\BCBL {}\ \BBA {} Gloor%
}{%
Elshendy%
\ \protect \BOthers {.}}{%
{\protect \APACyear {2018}}%
}]{%
elshendy2018using}
\APACinsertmetastar {%
elshendy2018using}%
\begin{APACrefauthors}%
Elshendy, M.%
, Colladon, A\BPBI F.%
, Battistoni, E.%
\BCBL {}\ \BBA {} Gloor, P\BPBI A.%
\end{APACrefauthors}%
\unskip\
\newblock
\APACrefYearMonthDay{2018}{}{}.
\newblock
{\BBOQ}\APACrefatitle {Using four different online media sources to forecast
  the crude oil price} {Using four different online media sources to forecast
  the crude oil price}.{\BBCQ}
\newblock
\APACjournalVolNumPages{Journal of Information Science}{44}{3}{408--421}.
\newblock
\begin{APACrefDOI} \doi{https://doi.org/10.1177/0165551517698298}
  \end{APACrefDOI}
\PrintBackRefs{\CurrentBib}

\bibitem [\protect \citeauthoryear {%
Emmert-Streib%
, Tripathi%
, Yli-Harja%
\BCBL {}\ \BBA {} Dehmer%
}{%
Emmert-Streib%
\ \protect \BOthers {.}}{%
{\protect \APACyear {2018}}%
}]{%
emmert2018understanding}
\APACinsertmetastar {%
emmert2018understanding}%
\begin{APACrefauthors}%
Emmert-Streib, F.%
, Tripathi, S.%
, Yli-Harja, O.%
\BCBL {}\ \BBA {} Dehmer, M.%
\end{APACrefauthors}%
\unskip\
\newblock
\APACrefYearMonthDay{2018}{}{}.
\newblock
{\BBOQ}\APACrefatitle {Understanding the world economy in terms of networks: A
  survey of data-based network science approaches on economic networks}
  {Understanding the world economy in terms of networks: A survey of data-based
  network science approaches on economic networks}.{\BBCQ}
\newblock
\APACjournalVolNumPages{Frontiers in Applied Mathematics and
  Statistics}{4}{}{37}.
\newblock
\begin{APACrefDOI} \doi{https://doi.org/10.3389/fams.2018.00037}
  \end{APACrefDOI}
\PrintBackRefs{\CurrentBib}

\bibitem [\protect \citeauthoryear {%
\APACcitebtitle {GDELT Project}}{%
\APACcitebtitle {GDELT Project}}{%
{\protect \APACyear {2015}}%
}]{%
leetaru_the_nodate}
\APACinsertmetastar {%
leetaru_the_nodate}%
\APACrefbtitle {GDELT Project.} {Gdelt project.}
\newblock
\APACrefYearMonthDay{2015}{}{}.
\newblock
\begin{APACrefURL} \url{https://www.gdeltproject.org/Accessed 15 May 2020}
  \end{APACrefURL}
\PrintBackRefs{\CurrentBib}

\bibitem [\protect \citeauthoryear {%
Ghalmane%
, Cherifi%
, Cherifi%
\BCBL {}\ \BBA {} El~Hassouni%
}{%
Ghalmane%
\ \protect \BOthers {.}}{%
{\protect \APACyear {2020}}%
}]{%
ghalmane2020extracting}
\APACinsertmetastar {%
ghalmane2020extracting}%
\begin{APACrefauthors}%
Ghalmane, Z.%
, Cherifi, C.%
, Cherifi, H.%
\BCBL {}\ \BBA {} El~Hassouni, M.%
\end{APACrefauthors}%
\unskip\
\newblock
\APACrefYearMonthDay{2020}{}{}.
\newblock
{\BBOQ}\APACrefatitle {Extracting backbones in weighted modular complex
  networks} {Extracting backbones in weighted modular complex networks}.{\BBCQ}
\newblock
\APACjournalVolNumPages{Scientific Reports}{10}{1}{1--18}.
\newblock
\begin{APACrefDOI} \doi{https://doi.org/10.1038/s41598-020-71876-0}
  \end{APACrefDOI}
\PrintBackRefs{\CurrentBib}

\bibitem [\protect \citeauthoryear {%
Giannone%
, Reichlin%
\BCBL {}\ \BBA {} Small%
}{%
Giannone%
\ \protect \BOthers {.}}{%
{\protect \APACyear {2008}}%
}]{%
giannone2008nowcasting}
\APACinsertmetastar {%
giannone2008nowcasting}%
\begin{APACrefauthors}%
Giannone, D.%
, Reichlin, L.%
\BCBL {}\ \BBA {} Small, D.%
\end{APACrefauthors}%
\unskip\
\newblock
\APACrefYearMonthDay{2008}{}{}.
\newblock
{\BBOQ}\APACrefatitle {Nowcasting: The real-time informational content of
  macroeconomic data} {Nowcasting: The real-time informational content of
  macroeconomic data}.{\BBCQ}
\newblock
\APACjournalVolNumPages{Journal of Monetary Economics}{55}{4}{665--676}.
\newblock
\begin{APACrefDOI} \doi{https://doi.org/10.1016/j.jmoneco.2008.05.010}
  \end{APACrefDOI}
\PrintBackRefs{\CurrentBib}

\bibitem [\protect \citeauthoryear {%
Girardi%
, Guardabascio%
\BCBL {}\ \BBA {} Ventura%
}{%
Girardi%
\ \protect \BOthers {.}}{%
{\protect \APACyear {2016}}%
}]{%
girardi2016factor}
\APACinsertmetastar {%
girardi2016factor}%
\begin{APACrefauthors}%
Girardi, A.%
, Guardabascio, B.%
\BCBL {}\ \BBA {} Ventura, M.%
\end{APACrefauthors}%
\unskip\
\newblock
\APACrefYearMonthDay{2016}{}{}.
\newblock
{\BBOQ}\APACrefatitle {Factor-Augmented Bridge Models (FABM) and Soft
  Indicators to Forecast Italian Industrial Production} {Factor-augmented
  bridge models (fabm) and soft indicators to forecast italian industrial
  production}.{\BBCQ}
\newblock
\APACjournalVolNumPages{Journal of Forecasting}{35}{6}{542--552}.
\newblock
\begin{APACrefDOI} \doi{https://doi.org/10.1002/for.2393} \end{APACrefDOI}
\PrintBackRefs{\CurrentBib}

\bibitem [\protect \citeauthoryear {%
Granger%
}{%
Granger%
}{%
{\protect \APACyear {1969}}%
}]{%
granger1969investigating}
\APACinsertmetastar {%
granger1969investigating}%
\begin{APACrefauthors}%
Granger, C\BPBI W.%
\end{APACrefauthors}%
\unskip\
\newblock
\APACrefYearMonthDay{1969}{}{}.
\newblock
{\BBOQ}\APACrefatitle {Investigating causal relations by econometric models and
  cross-spectral methods} {Investigating causal relations by econometric models
  and cross-spectral methods}.{\BBCQ}
\newblock
\APACjournalVolNumPages{Econometrica: journal of the Econometric
  Society}{}{}{424--438}.
\newblock
\begin{APACrefDOI} \doi{https://doi.org/10.2307/1912791} \end{APACrefDOI}
\PrintBackRefs{\CurrentBib}

\bibitem [\protect \citeauthoryear {%
Graves%
\ \BBA {} Schmidhuber%
}{%
Graves%
\ \BBA {} Schmidhuber%
}{%
{\protect \APACyear {2005}}%
}]{%
graves2005framewise}
\APACinsertmetastar {%
graves2005framewise}%
\begin{APACrefauthors}%
Graves, A.%
\BCBT {}\ \BBA {} Schmidhuber, J.%
\end{APACrefauthors}%
\unskip\
\newblock
\APACrefYearMonthDay{2005}{}{}.
\newblock
{\BBOQ}\APACrefatitle {Framewise phoneme classification with bidirectional LSTM
  networks} {Framewise phoneme classification with bidirectional lstm
  networks}.{\BBCQ}
\newblock
\BIn{} \APACrefbtitle {Proceedings. 2005 IEEE International Joint Conference on
  Neural Networks, 2005.} {Proceedings. 2005 ieee international joint
  conference on neural networks, 2005.}\ (\BVOL~4, \BPGS\ 2047--2052).
\PrintBackRefs{\CurrentBib}

\bibitem [\protect \citeauthoryear {%
Guo%
\ \BBA {} Vargo%
}{%
Guo%
\ \BBA {} Vargo%
}{%
{\protect \APACyear {2020}}%
}]{%
guo2020predictors}
\APACinsertmetastar {%
guo2020predictors}%
\begin{APACrefauthors}%
Guo, L.%
\BCBT {}\ \BBA {} Vargo, C\BPBI J.%
\end{APACrefauthors}%
\unskip\
\newblock
\APACrefYearMonthDay{2020}{}{}.
\newblock
{\BBOQ}\APACrefatitle {Predictors of International News Flow: Exploring a
  Networked Global Media System} {Predictors of international news flow:
  Exploring a networked global media system}.{\BBCQ}
\newblock
\APACjournalVolNumPages{Journal of Broadcasting \& Electronic
  Media}{64}{3}{418--437}.
\newblock
\begin{APACrefDOI} \doi{https://doi.org/10.1080/08838151.2020.1796391}
  \end{APACrefDOI}
\PrintBackRefs{\CurrentBib}

\bibitem [\protect \citeauthoryear {%
Harvey%
, Leybourne%
\BCBL {}\ \BBA {} Newbold%
}{%
Harvey%
\ \protect \BOthers {.}}{%
{\protect \APACyear {1997}}%
}]{%
harvey1997testing}
\APACinsertmetastar {%
harvey1997testing}%
\begin{APACrefauthors}%
Harvey, D.%
, Leybourne, S.%
\BCBL {}\ \BBA {} Newbold, P.%
\end{APACrefauthors}%
\unskip\
\newblock
\APACrefYearMonthDay{1997}{}{}.
\newblock
{\BBOQ}\APACrefatitle {Testing the equality of prediction mean squared errors}
  {Testing the equality of prediction mean squared errors}.{\BBCQ}
\newblock
\APACjournalVolNumPages{International Journal of forecasting}{13}{2}{281--291}.
\newblock
\begin{APACrefDOI} \doi{https://doi.org/10.1016/S0169-2070(96)00719-4}
  \end{APACrefDOI}
\PrintBackRefs{\CurrentBib}

\bibitem [\protect \citeauthoryear {%
Hochreiter%
\ \BBA {} Schmidhuber%
}{%
Hochreiter%
\ \BBA {} Schmidhuber%
}{%
{\protect \APACyear {1997}}%
}]{%
hochreiter1997long}
\APACinsertmetastar {%
hochreiter1997long}%
\begin{APACrefauthors}%
Hochreiter, S.%
\BCBT {}\ \BBA {} Schmidhuber, J.%
\end{APACrefauthors}%
\unskip\
\newblock
\APACrefYearMonthDay{1997}{}{}.
\newblock
{\BBOQ}\APACrefatitle {Long short-term memory} {Long short-term memory}.{\BBCQ}
\newblock
\APACjournalVolNumPages{Neural computation}{9}{8}{1735--1780}.
\newblock
\begin{APACrefDOI} \doi{https://doi.org/10.1162/neco.1997.9.8.1735}
  \end{APACrefDOI}
\PrintBackRefs{\CurrentBib}

\bibitem [\protect \citeauthoryear {%
Keynes%
}{%
Keynes%
}{%
{\protect \APACyear {2018}}%
}]{%
keynes2018general}
\APACinsertmetastar {%
keynes2018general}%
\begin{APACrefauthors}%
Keynes, J\BPBI M.%
\end{APACrefauthors}%
\unskip\
\newblock
\APACrefYear{2018}.
\newblock
\APACrefbtitle {The general theory of employment, interest, and money} {The
  general theory of employment, interest, and money}.
\newblock
\APACaddressPublisher{}{Springer}.
\PrintBackRefs{\CurrentBib}

\bibitem [\protect \citeauthoryear {%
King%
, Schneer%
\BCBL {}\ \BBA {} White%
}{%
King%
\ \protect \BOthers {.}}{%
{\protect \APACyear {2017}}%
}]{%
king2017news}
\APACinsertmetastar {%
king2017news}%
\begin{APACrefauthors}%
King, G.%
, Schneer, B.%
\BCBL {}\ \BBA {} White, A.%
\end{APACrefauthors}%
\unskip\
\newblock
\APACrefYearMonthDay{2017}{}{}.
\newblock
{\BBOQ}\APACrefatitle {How the news media activate public expression and
  influence national agendas} {How the news media activate public expression
  and influence national agendas}.{\BBCQ}
\newblock
\APACjournalVolNumPages{Science}{358}{6364}{776--780}.
\newblock
\begin{APACrefDOI} \doi{https://doi.org/10.1126/science.aao1100}
  \end{APACrefDOI}
\PrintBackRefs{\CurrentBib}

\bibitem [\protect \citeauthoryear {%
Leamer%
}{%
Leamer%
}{%
{\protect \APACyear {1985}}%
}]{%
leamer_1985}
\APACinsertmetastar {%
leamer_1985}%
\begin{APACrefauthors}%
Leamer, E\BPBI E.%
\end{APACrefauthors}%
\unskip\
\newblock
\APACrefYearMonthDay{1985}{}{}.
\newblock
{\BBOQ}\APACrefatitle {Self-Interpretation} {Self-interpretation}.{\BBCQ}
\newblock
\APACjournalVolNumPages{Economics and Philosophy}{1}{2}{295–302}.
\newblock
\begin{APACrefDOI} \doi{doi.org/10.1017/S0266267100002546} \end{APACrefDOI}
\PrintBackRefs{\CurrentBib}

\bibitem [\protect \citeauthoryear {%
Leetaru%
}{%
Leetaru%
}{%
{\protect \APACyear {2012}}%
}]{%
leetaru2012fulltext}
\APACinsertmetastar {%
leetaru2012fulltext}%
\begin{APACrefauthors}%
Leetaru, K\BPBI H.%
\end{APACrefauthors}%
\unskip\
\newblock
\APACrefYearMonthDay{2012}{}{}.
\newblock
{\BBOQ}\APACrefatitle {Fulltext geocoding versus spatial metadata for large
  text archives: Towards a geographically enriched Wikipedia} {Fulltext
  geocoding versus spatial metadata for large text archives: Towards a
  geographically enriched wikipedia}.{\BBCQ}
\newblock
\APACjournalVolNumPages{D-lib Magazine}{18}{9}{5}.
\newblock
\begin{APACrefDOI} \doi{https://doi.org/10.1045/september2012-leetaru}
  \end{APACrefDOI}
\PrintBackRefs{\CurrentBib}

\bibitem [\protect \citeauthoryear {%
Leetaru%
, Perkins%
\BCBL {}\ \BBA {} Rewerts%
}{%
Leetaru%
\ \protect \BOthers {.}}{%
{\protect \APACyear {2014}}%
}]{%
leetaru2014cultural}
\APACinsertmetastar {%
leetaru2014cultural}%
\begin{APACrefauthors}%
Leetaru, K\BPBI H.%
, Perkins, T.%
\BCBL {}\ \BBA {} Rewerts, C.%
\end{APACrefauthors}%
\unskip\
\newblock
\APACrefYearMonthDay{2014}{}{}.
\newblock
{\BBOQ}\APACrefatitle {Cultural computing at literature scale: encoding the
  cultural knowledge of tens of billions of words of academic literature}
  {Cultural computing at literature scale: encoding the cultural knowledge of
  tens of billions of words of academic literature}.{\BBCQ}
\newblock
\APACjournalVolNumPages{D-lib Magazine}{20}{9}{8}.
\newblock
\begin{APACrefDOI} \doi{https://doi.org/10.1045/september2014-leetaru}
  \end{APACrefDOI}
\PrintBackRefs{\CurrentBib}

\bibitem [\protect \citeauthoryear {%
Marcaccioli%
\ \BBA {} Livan%
}{%
Marcaccioli%
\ \BBA {} Livan%
}{%
{\protect \APACyear {2019}}%
}]{%
marcaccioli2019polya}
\APACinsertmetastar {%
marcaccioli2019polya}%
\begin{APACrefauthors}%
Marcaccioli, R.%
\BCBT {}\ \BBA {} Livan, G.%
\end{APACrefauthors}%
\unskip\
\newblock
\APACrefYearMonthDay{2019}{}{}.
\newblock
{\BBOQ}\APACrefatitle {A P{\'o}lya urn approach to information filtering in
  complex networks} {A p{\'o}lya urn approach to information filtering in
  complex networks}.{\BBCQ}
\newblock
\APACjournalVolNumPages{Nature communications}{10}{1}{1--10}.
\newblock
\begin{APACrefDOI} \doi{https://doi.org/10.1038/s41467-019-08667-3}
  \end{APACrefDOI}
\PrintBackRefs{\CurrentBib}

\bibitem [\protect \citeauthoryear {%
Matesanz~Gomez%
, Ferrari%
, Torgler%
\BCBL {}\ \BBA {} Ortega%
}{%
Matesanz~Gomez%
\ \protect \BOthers {.}}{%
{\protect \APACyear {2017}}%
}]{%
matesanz2017synchronization}
\APACinsertmetastar {%
matesanz2017synchronization}%
\begin{APACrefauthors}%
Matesanz~Gomez, D.%
, Ferrari, H\BPBI J.%
, Torgler, B.%
\BCBL {}\ \BBA {} Ortega, G\BPBI J.%
\end{APACrefauthors}%
\unskip\
\newblock
\APACrefYearMonthDay{2017}{}{}.
\newblock
{\BBOQ}\APACrefatitle {Synchronization and diversity in business cycles: a
  network analysis of the European Union} {Synchronization and diversity in
  business cycles: a network analysis of the european union}.{\BBCQ}
\newblock
\APACjournalVolNumPages{Applied Economics}{49}{10}{972--986}.
\newblock
\begin{APACrefDOI} \doi{https://doi.org/10.1080/00036846.2016.1210765}
  \end{APACrefDOI}
\PrintBackRefs{\CurrentBib}

\bibitem [\protect \citeauthoryear {%
McCracken%
\ \BBA {} Ng%
}{%
McCracken%
\ \BBA {} Ng%
}{%
{\protect \APACyear {2016}}%
}]{%
mccracken2016fred}
\APACinsertmetastar {%
mccracken2016fred}%
\begin{APACrefauthors}%
McCracken, M\BPBI W.%
\BCBT {}\ \BBA {} Ng, S.%
\end{APACrefauthors}%
\unskip\
\newblock
\APACrefYearMonthDay{2016}{}{}.
\newblock
{\BBOQ}\APACrefatitle {FRED-MD: A monthly database for macroeconomic research}
  {Fred-md: A monthly database for macroeconomic research}.{\BBCQ}
\newblock
\APACjournalVolNumPages{Journal of Business \& Economic
  Statistics}{34}{4}{574--589}.
\newblock
\begin{APACrefDOI} \doi{https://doi.org/10.1080/07350015.2015.1086655}
  \end{APACrefDOI}
\PrintBackRefs{\CurrentBib}

\bibitem [\protect \citeauthoryear {%
Nyman%
\ \protect \BOthers {.}}{%
Nyman%
\ \protect \BOthers {.}}{%
{\protect \APACyear {2018}}%
}]{%
nyman2018news}
\APACinsertmetastar {%
nyman2018news}%
\begin{APACrefauthors}%
Nyman, R.%
, Kapadia, S.%
, Tuckett, D.%
, Gregory, D.%
, Ormerod, P.%
\BCBL {}\ \BBA {} Smith, R.%
\end{APACrefauthors}%
\unskip\
\newblock
\APACrefYearMonthDay{2018}{}{}.
\newblock
\APACrefbtitle {News and narratives in financial systems: exploiting big data
  for systemic risk assessment.} {News and narratives in financial systems:
  exploiting big data for systemic risk assessment.}
\newblock
\begin{APACrefURL}
  \url{https://www.bankofengland.co.uk/working-paper/2018/news-and-narratives-in-financial-systems/Accessed
  30 October 2019} \end{APACrefURL}
\PrintBackRefs{\CurrentBib}

\bibitem [\protect \citeauthoryear {%
Piccardi%
\ \BBA {} Tajoli%
}{%
Piccardi%
\ \BBA {} Tajoli%
}{%
{\protect \APACyear {2018}}%
}]{%
piccardi2018complexity}
\APACinsertmetastar {%
piccardi2018complexity}%
\begin{APACrefauthors}%
Piccardi, C.%
\BCBT {}\ \BBA {} Tajoli, L.%
\end{APACrefauthors}%
\unskip\
\newblock
\APACrefYearMonthDay{2018}{}{}.
\newblock
{\BBOQ}\APACrefatitle {Complexity, centralization, and fragility in economic
  networks} {Complexity, centralization, and fragility in economic
  networks}.{\BBCQ}
\newblock
\APACjournalVolNumPages{PloS one}{13}{11}{e0208265}.
\newblock
\begin{APACrefDOI} \doi{https://doi.org/10.1371/journal.pone.0208265}
  \end{APACrefDOI}
\PrintBackRefs{\CurrentBib}

\bibitem [\protect \citeauthoryear {%
Schuster%
\ \BBA {} Paliwal%
}{%
Schuster%
\ \BBA {} Paliwal%
}{%
{\protect \APACyear {1997}}%
}]{%
schuster1997bidirectional}
\APACinsertmetastar {%
schuster1997bidirectional}%
\begin{APACrefauthors}%
Schuster, M.%
\BCBT {}\ \BBA {} Paliwal, K\BPBI K.%
\end{APACrefauthors}%
\unskip\
\newblock
\APACrefYearMonthDay{1997}{}{}.
\newblock
{\BBOQ}\APACrefatitle {Bidirectional recurrent neural networks} {Bidirectional
  recurrent neural networks}.{\BBCQ}
\newblock
\APACjournalVolNumPages{IEEE transactions on Signal
  Processing}{45}{11}{2673--2681}.
\newblock
\begin{APACrefDOI} \doi{https://doi.org/10.1109/78.650093} \end{APACrefDOI}
\PrintBackRefs{\CurrentBib}

\bibitem [\protect \citeauthoryear {%
Serrano%
, Bogun{\'a}%
\BCBL {}\ \BBA {} Vespignani%
}{%
Serrano%
\ \protect \BOthers {.}}{%
{\protect \APACyear {2009}}%
}]{%
serrano2009extracting}
\APACinsertmetastar {%
serrano2009extracting}%
\begin{APACrefauthors}%
Serrano, M\BPBI {\'A}.%
, Bogun{\'a}, M.%
\BCBL {}\ \BBA {} Vespignani, A.%
\end{APACrefauthors}%
\unskip\
\newblock
\APACrefYearMonthDay{2009}{}{}.
\newblock
{\BBOQ}\APACrefatitle {Extracting the multiscale backbone of complex weighted
  networks} {Extracting the multiscale backbone of complex weighted
  networks}.{\BBCQ}
\newblock
\APACjournalVolNumPages{Proceedings of the national academy of
  sciences}{106}{16}{6483--6488}.
\newblock
\begin{APACrefDOI} \doi{https://doi.org/10.1073/pnas.0808904106}
  \end{APACrefDOI}
\PrintBackRefs{\CurrentBib}

\bibitem [\protect \citeauthoryear {%
Shiller%
}{%
Shiller%
}{%
{\protect \APACyear {2017}}%
}]{%
shiller2017narrative}
\APACinsertmetastar {%
shiller2017narrative}%
\begin{APACrefauthors}%
Shiller, R\BPBI J.%
\end{APACrefauthors}%
\unskip\
\newblock
\APACrefYearMonthDay{2017}{}{}.
\newblock
{\BBOQ}\APACrefatitle {Narrative economics} {Narrative economics}.{\BBCQ}
\newblock
\APACjournalVolNumPages{American Economic Review}{107}{4}{967--1004}.
\newblock
\begin{APACrefDOI} \doi{https://doi.org/10.1257/aer.107.4.967} \end{APACrefDOI}
\PrintBackRefs{\CurrentBib}

\bibitem [\protect \citeauthoryear {%
Smets%
\ \BBA {} Wouters%
}{%
Smets%
\ \BBA {} Wouters%
}{%
{\protect \APACyear {2007}}%
}]{%
smets2007shocks}
\APACinsertmetastar {%
smets2007shocks}%
\begin{APACrefauthors}%
Smets, F.%
\BCBT {}\ \BBA {} Wouters, R.%
\end{APACrefauthors}%
\unskip\
\newblock
\APACrefYearMonthDay{2007}{}{}.
\newblock
{\BBOQ}\APACrefatitle {Shocks and frictions in US business cycles: A Bayesian
  DSGE approach} {Shocks and frictions in us business cycles: A bayesian dsge
  approach}.{\BBCQ}
\newblock
\APACjournalVolNumPages{American economic review}{97}{3}{586--606}.
\newblock
\begin{APACrefDOI} \doi{https://doi.org/10.1257/aer.97.3.586} \end{APACrefDOI}
\PrintBackRefs{\CurrentBib}

\bibitem [\protect \citeauthoryear {%
Stern%
, Livan%
\BCBL {}\ \BBA {} Smith%
}{%
Stern%
\ \protect \BOthers {.}}{%
{\protect \APACyear {2020}}%
}]{%
stern2020network}
\APACinsertmetastar {%
stern2020network}%
\begin{APACrefauthors}%
Stern, S.%
, Livan, G.%
\BCBL {}\ \BBA {} Smith, R\BPBI E.%
\end{APACrefauthors}%
\unskip\
\newblock
\APACrefYearMonthDay{2020}{}{}.
\newblock
{\BBOQ}\APACrefatitle {A network perspective on intermedia agenda-setting} {A
  network perspective on intermedia agenda-setting}.{\BBCQ}
\newblock
\APACjournalVolNumPages{arXiv preprint arXiv:2002.05971}{}{}{}.
\newblock
\begin{APACrefDOI} \doi{https://doi.org/10.1007/s41109-020-00272-4}
  \end{APACrefDOI}
\PrintBackRefs{\CurrentBib}

\bibitem [\protect \citeauthoryear {%
Stock%
\ \BBA {} Watson%
}{%
Stock%
\ \BBA {} Watson%
}{%
{\protect \APACyear {2001}}%
}]{%
stock2001vector}
\APACinsertmetastar {%
stock2001vector}%
\begin{APACrefauthors}%
Stock, J\BPBI H.%
\BCBT {}\ \BBA {} Watson, M\BPBI W.%
\end{APACrefauthors}%
\unskip\
\newblock
\APACrefYearMonthDay{2001}{}{}.
\newblock
{\BBOQ}\APACrefatitle {Vector autoregressions} {Vector autoregressions}.{\BBCQ}
\newblock
\APACjournalVolNumPages{Journal of Economic perspectives}{15}{4}{101--115}.
\newblock
\begin{APACrefDOI} \doi{https://doi.org/10.1257/jep.15.4.101} \end{APACrefDOI}
\PrintBackRefs{\CurrentBib}

\bibitem [\protect \citeauthoryear {%
Tantardini%
, Ieva%
, Tajoli%
\BCBL {}\ \BBA {} Piccardi%
}{%
Tantardini%
\ \protect \BOthers {.}}{%
{\protect \APACyear {2019}}%
}]{%
tandardini_2019_comparing}
\APACinsertmetastar {%
tandardini_2019_comparing}%
\begin{APACrefauthors}%
Tantardini, M.%
, Ieva, F.%
, Tajoli, L.%
\BCBL {}\ \BBA {} Piccardi, C.%
\end{APACrefauthors}%
\unskip\
\newblock
\APACrefYearMonthDay{2019}{11}{}.
\newblock
{\BBOQ}\APACrefatitle {Comparing methods for comparing networks} {Comparing
  methods for comparing networks}.{\BBCQ}
\newblock
\APACjournalVolNumPages{Scientific Reports}{9}{}{}.
\newblock
\begin{APACrefDOI} \doi{https://doi.org/10.1038/s41598-019-53708-y}
  \end{APACrefDOI}
\PrintBackRefs{\CurrentBib}

\bibitem [\protect \citeauthoryear {%
Temizsoy%
, Iori%
\BCBL {}\ \BBA {} Montes-Rojas%
}{%
Temizsoy%
\ \protect \BOthers {.}}{%
{\protect \APACyear {2016}}%
}]{%
temizsoy2016network}
\APACinsertmetastar {%
temizsoy2016network}%
\begin{APACrefauthors}%
Temizsoy, A.%
, Iori, G.%
\BCBL {}\ \BBA {} Montes-Rojas, G.%
\end{APACrefauthors}%
\unskip\
\newblock
\APACrefYearMonthDay{2016}{}{}.
\newblock
{\BBOQ}\APACrefatitle {Network Centrality and Funding Rates in the e-MID
  Interbank Market (16/08)} {Network centrality and funding rates in the e-mid
  interbank market (16/08)}.{\BBCQ}
\newblock
\APACjournalVolNumPages{London, UK: Department of Economics, City, University
  of London. This is the published version of the paper. This version of the
  publication may differ from the final published version}{}{}{}.
\newblock
\begin{APACrefDOI} \doi{https://doi.org/10.1016/j.jfs.2016.11.003}
  \end{APACrefDOI}
\PrintBackRefs{\CurrentBib}

\bibitem [\protect \citeauthoryear {%
Tilly%
, Ebner%
\BCBL {}\ \BBA {} Livan%
}{%
Tilly%
\ \protect \BOthers {.}}{%
{\protect \APACyear {2021}}%
}]{%
tilly2021forecasting}
\APACinsertmetastar {%
tilly2021forecasting}%
\begin{APACrefauthors}%
Tilly, S.%
, Ebner, M.%
\BCBL {}\ \BBA {} Livan, G.%
\end{APACrefauthors}%
\unskip\
\newblock
\APACrefYearMonthDay{2021}{}{}.
\newblock
{\BBOQ}\APACrefatitle {Macroeconomic forecasting through news, emotions and
  narrative} {Macroeconomic forecasting through news, emotions and
  narrative}.{\BBCQ}
\newblock
\APACjournalVolNumPages{Expert Systems and Applications}{}{}{}.
\newblock
\begin{APACrefDOI} \doi{https://doi.org/10.1016/j.eswa.2021.114760}
  \end{APACrefDOI}
\PrintBackRefs{\CurrentBib}

\bibitem [\protect \citeauthoryear {%
Tobias%
}{%
Tobias%
}{%
{\protect \APACyear {1995}}%
}]{%
tobias1995introduction}
\APACinsertmetastar {%
tobias1995introduction}%
\begin{APACrefauthors}%
Tobias, R\BPBI D.%
\end{APACrefauthors}%
\unskip\
\newblock
\APACrefYearMonthDay{1995}{}{}.
\newblock
{\BBOQ}\APACrefatitle {An introduction to partial least squares regression} {An
  introduction to partial least squares regression}.{\BBCQ}
\newblock
\BIn{} \APACrefbtitle {Proceedings of the twentieth annual SAS users group
  international conference} {Proceedings of the twentieth annual sas users
  group international conference}\ (\BVOL~20).
\PrintBackRefs{\CurrentBib}

\bibitem [\protect \citeauthoryear {%
Tuckett%
, Ormerod%
, Smith%
\BCBL {}\ \BBA {} Nyman%
}{%
Tuckett%
\ \protect \BOthers {.}}{%
{\protect \APACyear {2014}}%
}]{%
tuckett2014bringing}
\APACinsertmetastar {%
tuckett2014bringing}%
\begin{APACrefauthors}%
Tuckett, D.%
, Ormerod, P.%
, Smith, R.%
\BCBL {}\ \BBA {} Nyman, R.%
\end{APACrefauthors}%
\unskip\
\newblock
\APACrefYearMonthDay{2014}{}{}.
\newblock
{\BBOQ}\APACrefatitle {Bringing Social-Psychological Variables into Economic
  Modelling: Uncertainty, Animal Spirits and the Recovery from the Great
  Recession} {Bringing social-psychological variables into economic modelling:
  Uncertainty, animal spirits and the recovery from the great
  recession}.{\BBCQ}
\newblock
\APACjournalVolNumPages{Economic Growth eJournal}{}{}{}.
\newblock
\begin{APACrefDOI} \doi{https://doi.org/10.2139/ssrn.2408155} \end{APACrefDOI}
\PrintBackRefs{\CurrentBib}

\bibitem [\protect \citeauthoryear {%
Tumminello%
, Micciche%
, Lillo%
, Piilo%
\BCBL {}\ \BBA {} Mantegna%
}{%
Tumminello%
\ \protect \BOthers {.}}{%
{\protect \APACyear {2011}}%
}]{%
tumminello2011statistically}
\APACinsertmetastar {%
tumminello2011statistically}%
\begin{APACrefauthors}%
Tumminello, M.%
, Micciche, S.%
, Lillo, F.%
, Piilo, J.%
\BCBL {}\ \BBA {} Mantegna, R\BPBI N.%
\end{APACrefauthors}%
\unskip\
\newblock
\APACrefYearMonthDay{2011}{}{}.
\newblock
{\BBOQ}\APACrefatitle {Statistically validated networks in bipartite complex
  systems} {Statistically validated networks in bipartite complex
  systems}.{\BBCQ}
\newblock
\APACjournalVolNumPages{PloS one}{6}{3}{e17994}.
\newblock
\begin{APACrefDOI} \doi{https://doi.org/10.1371/journal.pone.0017994}
  \end{APACrefDOI}
\PrintBackRefs{\CurrentBib}

\bibitem [\protect \citeauthoryear {%
Van~Eyden%
, Difeto%
, Gupta%
\BCBL {}\ \BBA {} Wohar%
}{%
Van~Eyden%
\ \protect \BOthers {.}}{%
{\protect \APACyear {2019}}%
}]{%
van2019oil}
\APACinsertmetastar {%
van2019oil}%
\begin{APACrefauthors}%
Van~Eyden, R.%
, Difeto, M.%
, Gupta, R.%
\BCBL {}\ \BBA {} Wohar, M\BPBI E.%
\end{APACrefauthors}%
\unskip\
\newblock
\APACrefYearMonthDay{2019}{}{}.
\newblock
{\BBOQ}\APACrefatitle {Oil price volatility and economic growth: Evidence from
  advanced economies using more than a century’s data} {Oil price volatility
  and economic growth: Evidence from advanced economies using more than a
  century’s data}.{\BBCQ}
\newblock
\APACjournalVolNumPages{Applied Energy}{233}{}{612--621}.
\newblock
\begin{APACrefDOI} \doi{https://doi.org/10.1016/j.apenergy.2018.10.049}
  \end{APACrefDOI}
\PrintBackRefs{\CurrentBib}

\bibitem [\protect \citeauthoryear {%
Yang%
, Pang%
\BCBL {}\ \BBA {} Huang%
}{%
Yang%
\ \protect \BOthers {.}}{%
{\protect \APACyear {2020}}%
}]{%
yang2020knowledge}
\APACinsertmetastar {%
yang2020knowledge}%
\begin{APACrefauthors}%
Yang, Y.%
, Pang, Y.%
\BCBL {}\ \BBA {} Huang, G.%
\end{APACrefauthors}%
\unskip\
\newblock
\APACrefYearMonthDay{2020}{}{}.
\newblock
{\BBOQ}\APACrefatitle {The Knowledge Graph for Macroeconomic Analysis with
  Alternative Big Data} {The knowledge graph for macroeconomic analysis with
  alternative big data}.{\BBCQ}
\newblock
\APACjournalVolNumPages{arXiv preprint arXiv:2010.05172}{}{}{}.
\PrintBackRefs{\CurrentBib}

\end{thebibliography}

\end{document}